\documentclass[a4paper, 12pt]{article}

\usepackage{amssymb}
\usepackage{amsmath}
\usepackage{graphicx}
\usepackage[switch]{lineno}

\begin{document}

\title{Hidden drifts in turbulence}

\author{M. Vlad \footnote{madalina.vlad@inflpr.ro} \ and F. Spineanu \\
National Institute of Laser, Plasma and Radiation Physics \\
 Atomistilor 409 Magurele 077125 Romania}



\maketitle

\abstract{
The paper defines and discusses the concept of hidden drifts in
two-dimensional turbulence. These are ordered components of the trajectories
that average to zero and do not produce direct transport. Their effects
appear in the evolution of the turbulence as a special type of fluxes, which
consist of average motion of positive and negative fluctuations in opposite
directions. We show that these fluxes have important nonlinear effects in
turbulent fluids and in confined plasmas. In the first case, they determine
the increase of the large scale vorticity and velocity at the expense of the
small scale fluctuations by a process of separation of the vorticity
fluctuations according to their sign. In the second case, they provide a
mechanism for zonal flow generation and a vorticity flux that influences the
sheared rotation of the plasma.

\bigskip

Pacs: 02.50.Ey (Stochastic processes); 52.35.Ra (plasma turbulence); 47.27.-i (turbulent flows).
}


\section{Introduction}

Turbulence is of fundamental importance in fluid mechanics, plasma physics,
astrophysics, atmosphere and ocean sciences. It is a complex nonlinear
process that mixes disorder and order \cite{Falkovich}-\cite{Fauve}.
Quasi-coherence or order appears at the basic level of tracer (or test
particle) trajectories in the case of smooth velocity fields that have
finite correlation lengths $\lambda .$ It lasts for a short time of the
order of the time of flight $\tau _{fl}=\lambda /V,$ where $V$ is the
amplitude of the stochastic velocity. In two-dimensional (2D)  incompressible turbulence, long time coherence can appear essentially because the
trajectories are trapped in the correlated zone. Tracer trajectories are
random sequences of trapping or eddying events and long jumps. The latter
are random, while the motion during the trapping events has a high degree of
order, which strongly modifies the diffusion process \cite{kraichnan}-\cite%
{VS13}.

The order of the trajectories determines much amplified and complicated
effects on turbulence fields due to the nonlinear constraints that
characterize of the self-consistent evolution.

The turbulence that is dominantly 2D has a self-organizing
character \cite{Montgomery}-\cite{Weisse}, which consists of the generation
of quasi-coherent structures and flows. In addition to that, we have found
more subtle effects, namely the hidden drifts (HDs).

Here we define the HDs and analyze some of their effects on turbulence
evolution. The HDs are discovered in the statistics of tracers trajectories,
but their main influence is on turbulence evolution rather than on tracer
transport. Essentially, the HDs are two opposite average velocities that
compensate one another. We show that the HDs are the origin of strange
turbulent fluxes (STFs), which consist of an advection process that depends
on the sign of the fluctuations. The sign of the advected fluctuations is
associated to the sign of the HD, so that the positive and the negative
fluctuations move in opposite directions and generate fluxes. The STFs
contribute to the understanding of important nonlinear processes in fluid
and plasma turbulence.

The HDs and the STFs are fundamental processes in all physical domains where 2D turbulence is important. We discuss here two representative
systems: turbulence relaxation in ideal fluids \cite{Montgomery}-\cite%
{Proven} and drift turbulence in magnetically confined plasmas \cite{K02}.
We show that the STFs provide mechanism for the separation of the vorticity
according to its sign and for the inverse energy cascade in fluids \cite%
{Weisse}-\cite{Dubin}. The effects of the STFs are different in turbulent
plasmas, where they contribute to the generation of zonal flow modes \cite%
{Tzf}-\cite{Plunk} that can lead to improved confinement (\cite{Estrada}-%
\cite{Schmitz}).

These examples show that the effects of the HDs depend on the nature of the
nonlinearity and on the physical significance of the advected fields.

\section{Hidden drifts}

We start from the statistics of tracer trajectories in 2D
stochastic velocity fields obtained from

\begin{equation}
\frac{d\mathbf{x}}{dt}=\mathbf{v}(\mathbf{x,}t)+V_{d}\mathbf{e}_{2},~\ \ 
\mathbf{v}(\mathbf{x,}t)=-\mathbf{\nabla }\phi (\mathbf{x,}t)\times \mathbf{e%
}_{3},  \label{eqm}
\end{equation}%
where $\mathbf{e}_{1},$ $\mathbf{e}_{2}$ are the unit vectors in the plane
of the motion $\mathbf{x=(}x_{1},x_{2}),$ $\mathbf{e}_{3}$ is perpendicular
on this plane and $V_{d}$ is a constant average velocity. The initial
condition is $\mathbf{x=0}$ at $t=0.$ The stochastic velocity field $\mathbf{%
v}(\mathbf{x,}t),$ determined by the potential (or stream function) $\phi (%
\mathbf{x,}t),$\ has zero divergence. The potential is a stationary and
homogeneous stochastic Gaussian field with zero average and with the
Eulerian correlation (EC)%
\begin{equation}
E(\mathbf{x,}t)=\left\langle \phi (\mathbf{0,}0)~\phi (\mathbf{x,}%
t)\right\rangle  \label{EC}
\end{equation}%
modelled according to the characteristics of the physical system. Usually, $%
E(\mathbf{x,}t)$ decays from the maximum $E(\mathbf{0,}0)=\Phi ^{2}$ to zero
at the characteristic distances $\lambda _{1},$ $\lambda _{2}$\ and/or time $%
\tau _{d}.$ It defines the main parameters of the stochastic potential: the
amplitude $\Phi ,$ the decorrelation lengths $\lambda _{i}$ and the
decorrelation time $\tau _{d}$. The Hamiltonian structure of equation (\ref%
{eqm}) is the origin of the order that characterizes the 2D
incompressible turbulence. The velocity is tangent to the contour lines of
the total potential $\phi _{t}(\mathbf{x,}t)=\phi (\mathbf{x,}t)+x_{1}V_{d}$%
\ at any moment, and, in the case of time independent potentials, the
trajectories remain on these lines. They reflect the space structure of the
potential.\ \ 

The HDs are ordered displacements that average to zero and do not drive
flows. They appear in the presence of an average velocity $\mathbf{V}_{d}$
and they are perpendicular on $\mathbf{V}_{d}$. The HDs are found by
analyzing the average displacements conditioned by the initial value of the
potential $\left\langle x_{i}(t)\right\rangle _{\phi ^{0}},$ where $\phi
^{0}=\phi (\mathbf{0,}0)$.\ \ 

These conditional displacements and the other statistical characteristics of
the trajectories are determined using the decorrelation trajectory method
(DTM, \cite{V98}-\cite{VS17}, \cite{VS04}). This is a semi-analytic approach
that is in agreement with the statistical consequences of the invariance of
the potential. The main idea of this method is to determine the Lagrangian
averages not on the whole set of trajectories, but rather to group together
trajectories that are similar, to calculate their average and then to
evaluate averages of these averages. Similar trajectories are obtained by
imposing supplementary initial conditions, namely the values of the
potential and of the velocity at the starting point of the trajectory $\phi
^{0}=\phi (\mathbf{0,}0),\mathbf{v}^{0}=\mathbf{v}(\mathbf{0,}0).$ They
define a set of subensembles $S$ of the realizations of the potential.
Conditional averages lead to space-time dependent average potential $\Phi
^{S}(\mathbf{x,}t)$ and velocity $\mathbf{V}^{S}(\mathbf{x,}t)$ in each
subensemble $S,$ and to a decorrelation trajectory (DT) $\mathbf{X}(t;\phi
^{0},\mathbf{v}^{0})$. The DTs are the main ingredient of DTM. They are
smooth, simple trajectories determined from an equation with the same
structure as equation (\ref{eqm}), but with the stochastic potential
replaced by $\Phi ^{S}(\mathbf{x,}t)$ 
\begin{equation}
\frac{d\mathbf{X}}{dt}=\mathbf{v}(\mathbf{X,}t)+V_{d}\mathbf{e}_{2},~\ \ 
\mathbf{v}(\mathbf{X,}t)=-\mathbf{\nabla }\Phi ^{S}(\mathbf{X,}t)\times 
\mathbf{e}_{3}.  \label{eqDT}
\end{equation}%
The subensemble average potential $\Phi ^{S}(\mathbf{x,}t)$ is determined by
the EC of $\phi (\mathbf{x,}t)$

\begin{equation}
\Phi ^{S}(\mathbf{x,}t)=\phi ^{0}\frac{E(\mathbf{x,}t)}{E(\mathbf{0,}0)}%
-v_{1}^{0}\frac{E_{2}(\mathbf{x,}t)}{E_{22}(\mathbf{0,}0)}+v_{2}^{0}\frac{%
E_{1}(\mathbf{x,}t)}{E_{11}(\mathbf{0,}0)},  \label{FiS}
\end{equation}%
where $E_{ij}$ are derivatives of the type $E_{i}(\mathbf{x,}t)=\partial E(%
\mathbf{x,}t)/\partial x_{i}.$ The DTs represent the average evolution of
the particles through the correlated zone of the potential and describe the
decorrelation process.

The statistics of the trajectories is represented by averages along the DTs,
weighted by the Gaussian probability of the subensembles, $P\left( \phi
^{0})P(v_{1}^{0},v_{2}^{0}\right) .$ The DTM essentially determines the
correlations of the trajectories with the quantities that define the
subensembles. Rather complex transport processes could be analyzed with this
semi-analytical method in magnetically confined \cite{Jenko}-\cite{VS16} and
in space \cite{VSApJ}-\cite{VSCApJ} plasmas. It was shown that the DTM
yields clear physical images of the nonlinear stochastic advection and
reasonably good quantitative results.

The average displacements conditioned by the initial value of the potential $%
\left\langle x_{i}(t)\right\rangle _{\phi ^{0}}$ are evaluated in the frame
of the DTM by

\begin{equation}
\left\langle x_{i}(t)\right\rangle _{\phi ^{0}}=\int_{-\infty }^{\infty
}dv_{1}^{0}dv_{2}^{0}X_{i}(t;\phi ^{0},v_{i}^{0})P\left(
v_{1}^{0},v_{2}^{0}\right) .  \label{x-fi}
\end{equation}%
They are zero at any time when $V_{d}=0,$ but finite values of the component
along $\mathbf{e}_{1}$\ axis, $\left\langle x_{1}(t)\right\rangle _{\phi
^{0}},$\ yield in the presence of an average velocity $V_{d}$\ directed
along $\mathbf{e}_{2}$ axis.\ \ \ \qquad

The most important property of these average displacements is that they have
the same sign as the initial potential $\phi ^{0}$ at any time, as seen in
Figure 1a. The average displacements conditioned by the sign of the initial
potential are%
\begin{eqnarray}
\left\langle x_{1}(t)\right\rangle _{+} &=&\int_{0}^{\infty }d\phi
^{0}\left\langle x_{1}(t)\right\rangle _{\phi ^{0}}P(\phi ^{0}),\ \ 
\label{x-sign} \\
\left\langle x_{1}(t)\right\rangle _{-} &=&\int_{-\infty }^{0}d\phi
^{0}\left\langle x_{1}(t)\right\rangle _{\phi ^{0}}P(\phi ^{0}),  \notag
\end{eqnarray}%
where $P(\phi ^{0})$\ is the probability of $\phi ^{0}.$ They are, as seen
in Figure 1b, time-dependent functions that saturate. The sum $\left\langle
x_{i}(t)\right\rangle _{+}+\left\langle x_{i}(t)\right\rangle _{-}=0$, which
shows that there is no average motion.

Any process of decorrelation of the trajectories from the contour lines of
the potential leads to conditional average velocities. The trajectory $%
\mathbf{x(}t)$ for $t\gg \tau _{d}$ consists of a time sequence of segments
of duration $\tau _{d}$ that are statistically independent. The conditional
displacements (\ref{x-sign}) during the characteristic decorrelation time $%
\tau _{d}$\ determine the step length of the ordered motion and a pair of
average velocities

\begin{equation}
V_{+}=\frac{\left\langle x_{1}(\tau _{d})\right\rangle _{+}}{\tau _{d}},\
V_{-}=\frac{\left\langle x_{1}(\tau _{d})\right\rangle _{-}}{\tau _{d}}.
\label{Vpm}
\end{equation}%
They have opposite directions and compensate one another $V_{+}+V_{-}=0.$
These are the hidden drifts (HDs).

Thus, the tracer stochastic motion described by Eq. (\ref{eqm}) includes
ordered components, which determine the HDs, a pair of symmetrical
velocities perpendicular to the average velocity $V_{d}\mathbf{e}_{2}.$ The
characteristics of the motion that generate the HDs are presented in Figure
1. It demonstrates that the conditional average displacements $\left\langle
x_{i}(t)\right\rangle _{\phi ^{0}}$ (Eq. (\ref{x-fi})) have the sign of the
initial potential $\phi ^{0}$ at any time (Figure 1a). They lead, as seen in
Figure 1b, to perfectly symmetrical average displacements $\left\langle
x_{1}(t)\right\rangle _{+},$ $\left\langle x_{1}(t)\right\rangle _{-}$
(defined in Eq. (\ref{x-sign})), which are the origin of the HDs (\ref{Vpm}).


\begin{figure}[tbp]
\centerline{\includegraphics[height=4.2cm]{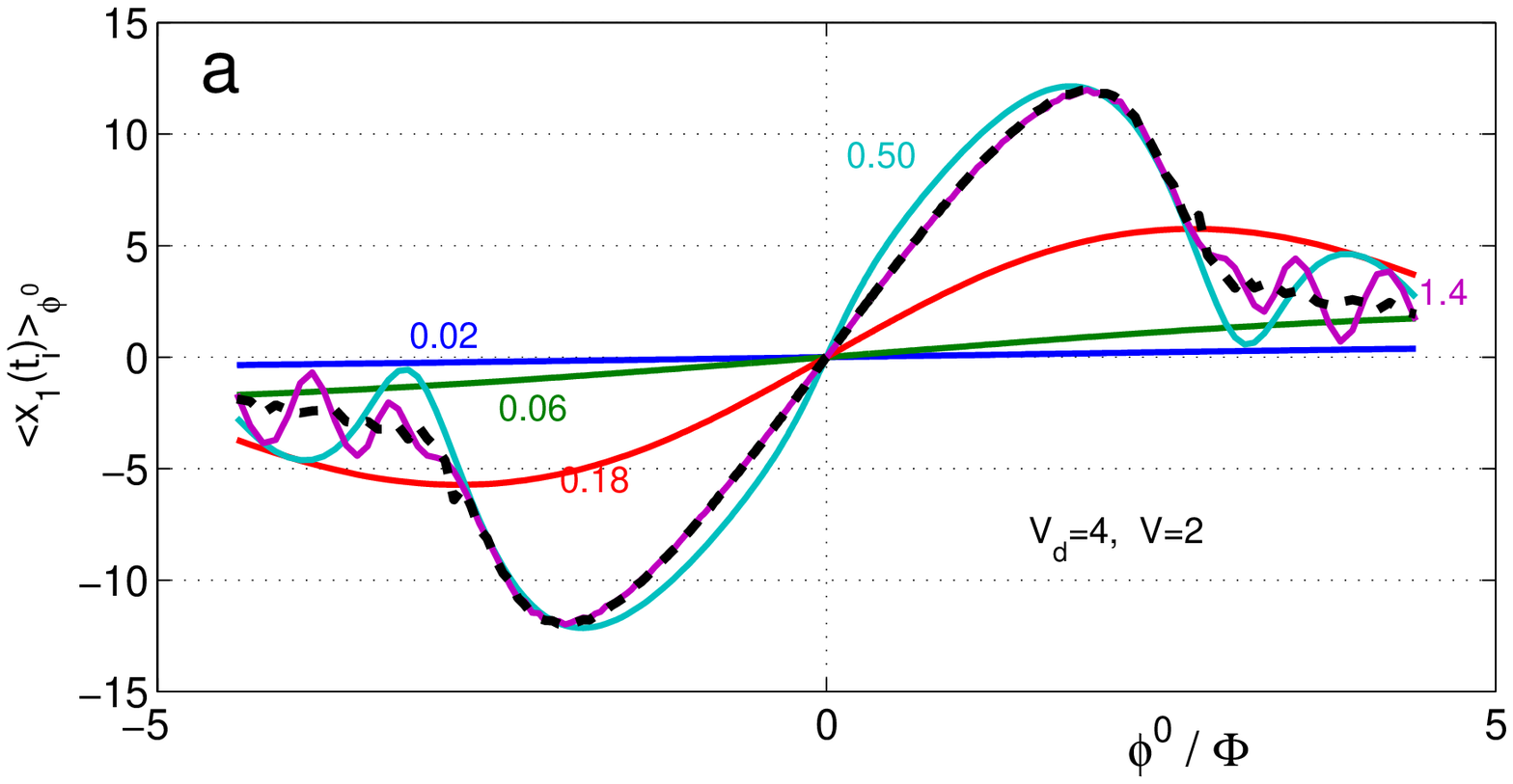}} \centerline{%
\includegraphics[height=4.2cm]{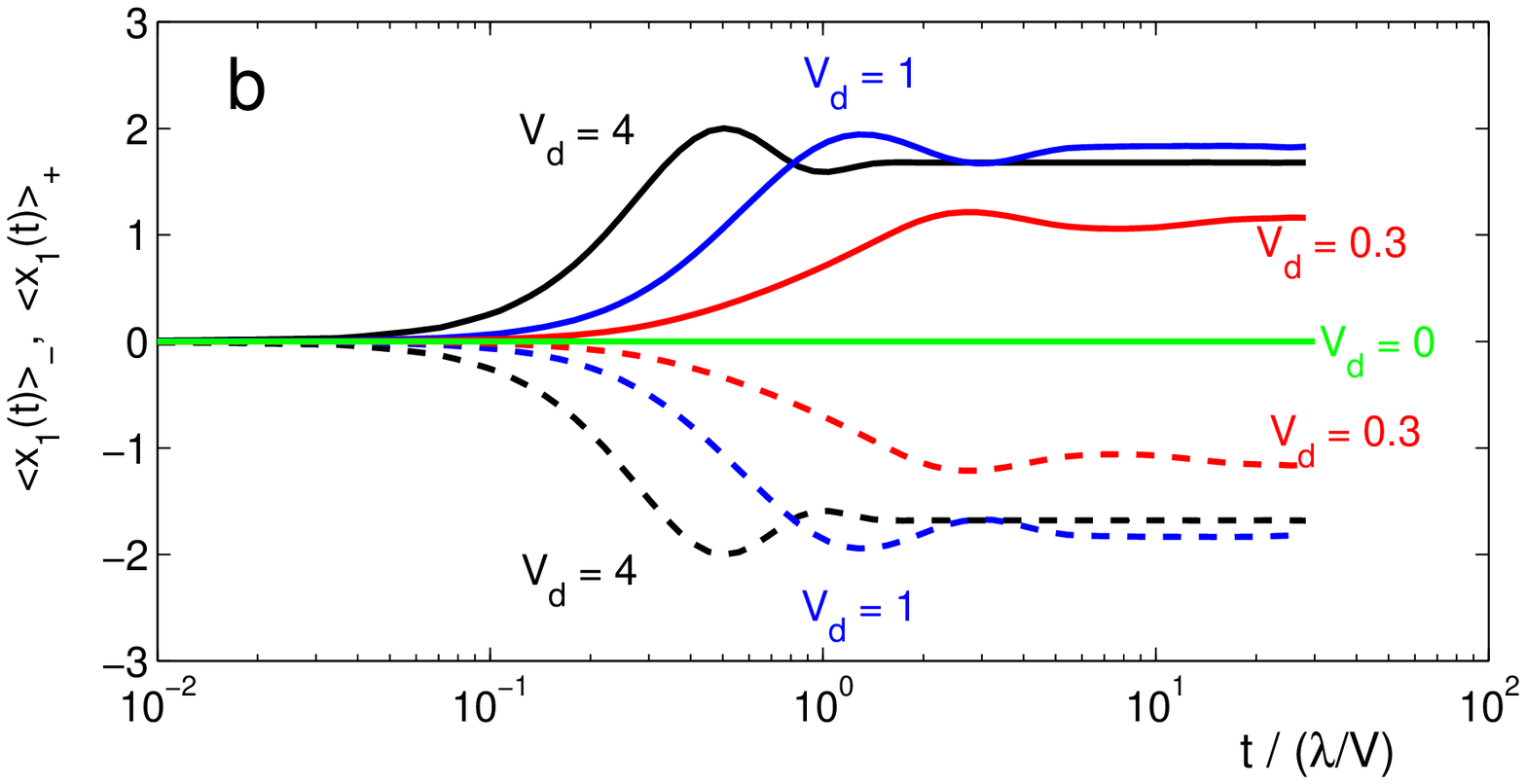}} 
\caption{The ordered motion that produces the HDs. a) $\left\langle
x_{1}(t)\right\rangle _{\protect\phi^{0}}$ as functions of $\protect\phi %
^{0} $ at the time moments $t_i$ that label the curves and at saturation
(dashed line). b) The conditional average displacements $\left\langle
x_{1}(t)\right\rangle _{-}$ (dashed lines) and $\left\langle
x_{1}(t)\right\rangle _{+}$ (continuous lines) as functions of time for the
values of $V_{d}$ that label the curves.}
\label{Figure 1}
\end{figure}

The physical explanation of the HDs is based on the modification of the
structure of the contour lines of the potential and on the perturbation of
the statistics of the velocity along these lines. Both effects are produced
by the average velocity $V_{d}\mathbf{e}_{2}.$

The average velocity determines an effective potential $\phi _{t}(\mathbf{x,}%
t)=\phi (\mathbf{x,}t)+x_{1}V_{d},$ which has strips of open contour lines
that extend along the average velocity, while oscillating in the
perpendicular direction. For small values of $V_{d}$ ($V_{d}<V),$ closed
field lines exist between the strips, but they are stretched along the average velocity and organized in pairs of apposite signs. 

In the closed potential cells, the velocity along the contour lines is
statistically non-homogeneous due to $V_{d}\mathbf{e}_{2},$ which leads to
total velocity $V_{t}$ that is enhanced on a side of the cells and\ reduced
on the opposite one.\ The trajectories are concentrated on the zones with
small $V_{t},$ which leads to an average displacement that has the sign of
the potential. The open strips also contribute to the conditional
displacements (\ref{x-sign}), but this is due to a different reason. The
invariance of the potential along an open trajectory with initial $\phi
^{0}, $ $\phi (\mathbf{x}(t))+x_{1}(t)V_{d}=\phi ^{0},$ shows that the
average displacement has the sign of $\phi ^{0},$ because the average
potential on these trajectories is zero.

Thus, the HDs reflect the order of the contour lines of the total potential.

We note that the ordered motion as symmetrical positive and negative
velocities that compensate each other is not forbidden by the
zero-divergence condition, which prohibits average Lagrangian velocities
along the $\mathbf{e}_{1}$ axis for the trajectories described by Eq. (\ref%
{eqm}).

The HDs do not have direct influence on transport. They do not drive a
direct transport (average velocity) and the contributions of the ordered
steps are implicitly included in the mean square displacements that define
the diffusion coefficients. However, they represent a reservoir for direct
transport, because perturbations produced by other components of the motion
that determine a weak compressibility of the velocity field can perturb the
equilibrium of the HDs leading to an average velocity of the test particles.
A first process of this type was recently found in the study of the
transport of the heavy impurities in turbulent plasmas \cite{VS2018}. The
perturbation is provided by the polarization drift, and it leads to direct
transport. The effect of the HDs is rather strong in this case since the
convective flux is comparable or larger than the diffusive transport for a
wide range of parameters of the transport model.

However, the most important effects of the HDs are connected to the special
turbulent fluxes described in the next sections.

\section{Strange turbulent fluxes}

The ordered displacements $\left\langle x_{1}(t)\right\rangle _{\phi ^{0}}$
that have the same sign as the initial potential lead to the correlation of
the displacements with the potential%
\begin{equation}
\left\langle \phi (\mathbf{0},0)~x_{1}(t)\right\rangle =\int_{-\infty
}^{\infty }d\phi ^{0}\phi ^{0}\left\langle x_{1}(t)\right\rangle _{\phi
^{0}}P(\phi ^{0}).  \label{xfi}
\end{equation}%
\ 

A similar correlation $\left\langle \omega (\mathbf{0},0)~x_{1}(t)\right%
\rangle $ exists between the displacement and the vorticity $\omega (\mathbf{%
x},t)=\bigtriangleup \phi (\mathbf{x},t).$ It is generated by the average
displacements conditioned by the initial vorticity $\left\langle
x_{1}(t)\right\rangle _{\omega ^{0}},$ where $\omega ^{0}=\omega (\mathbf{0,}%
0).$\ These averages are determined by extending the DTM. A supplementary
condition $\omega (\mathbf{0},0)=\omega ^{0}$ is introduced in the
definition of the subensembles. This determines a new term in the
subensemble average potential $\Phi ^{S}(\mathbf{X,}t)$ (\ref{FiS}), $\Phi
^{\omega }=\omega ^{0}\bigtriangleup ^{2}E(\mathbf{X},t)/\Omega ^{2},$ where 
$\Omega ^{2}=\bigtriangleup ^{2}E(\mathbf{0},0)$ is the amplitude of
vorticity fluctuations. As a result, the shapes of the DTs are strongly
changed and the probability of the subensembles is also modified due to the
correlation (\ref{EC_fiom}). However, this extended DTM leads to diffusion
coefficients that are not much modified.

The properties of $\left\langle x_{1}(t)\right\rangle _{\omega ^{0}}$\ are
similar to those of $\left\langle x_{1}(t)\right\rangle _{\phi ^{0}},$\ with
the difference that $\left\langle x_{1}(t)\right\rangle _{\omega ^{0}}$\ has
the sign opposite to the sign of $\omega ^{0}.$\ This difference is the
result of the negative correlation of $\phi $ and $\omega $\ 

\begin{equation}
\left\langle \phi (\mathbf{x,}t)\omega (\mathbf{x,}t)\right\rangle
=\bigtriangleup E(\mathbf{0},0)<0.  \label{EC_fiom}
\end{equation}

In the presence of a decorrelation process (at finite $\tau _{d}),$ the
correlations of the trajectories determine correlations of the Lagragian
velocity with the initial potential and vorticity, respectively\ \ 

\begin{eqnarray}
C_{\phi } &=&\left\langle \phi (\mathbf{0},0)~v_{x}(\mathbf{x}%
(t),t)\right\rangle \rightarrow \frac{\left\langle \phi (\mathbf{0}%
,0)~x_{1}(\tau _{d})\right\rangle }{\tau _{d}},  \label{Corfi} \\
C_{\omega } &=&\left\langle \omega (\mathbf{0},0)~v_{x}(\mathbf{x}%
(t),t)\right\rangle \rightarrow \frac{\left\langle \omega (\mathbf{0}%
,0)~x_{1}(\tau _{d})\right\rangle }{\tau _{d}}.  \label{Corvort}
\end{eqnarray}%
They are time dependent functions that saturate if $\tau _{d}$ is finite,
and decay to zero in static potentials.

The Lagrangian correlations $C_{\phi }$ and $C_{\omega }$\ depend on $V_{d}$%
, on the amplitude $V$ of the stochastic velocity and on the decorrelation
time $\tau _{d}.$\ Their dependence on these parameters is similar, except
for the signs, which are opposite. Both are anti-symmetrical functions of $%
V_{d}$\ that increase with $V$\ and decay at small and large $\tau _{d}.$
Typical results are presented in Figure 2. The anti-symmetrical dependence
on $V_{d}$\ is shown in Figure 2a, where $C_{\omega }$\ is represented for
several values of $\tau _{d}.$\ The dependence on $\tau _{d}$\ and $V$ is
shown in Figure 2b for $C_{\phi }.$\ \ 


\begin{figure}[tbp]
\centerline{\includegraphics[height=4.2cm]{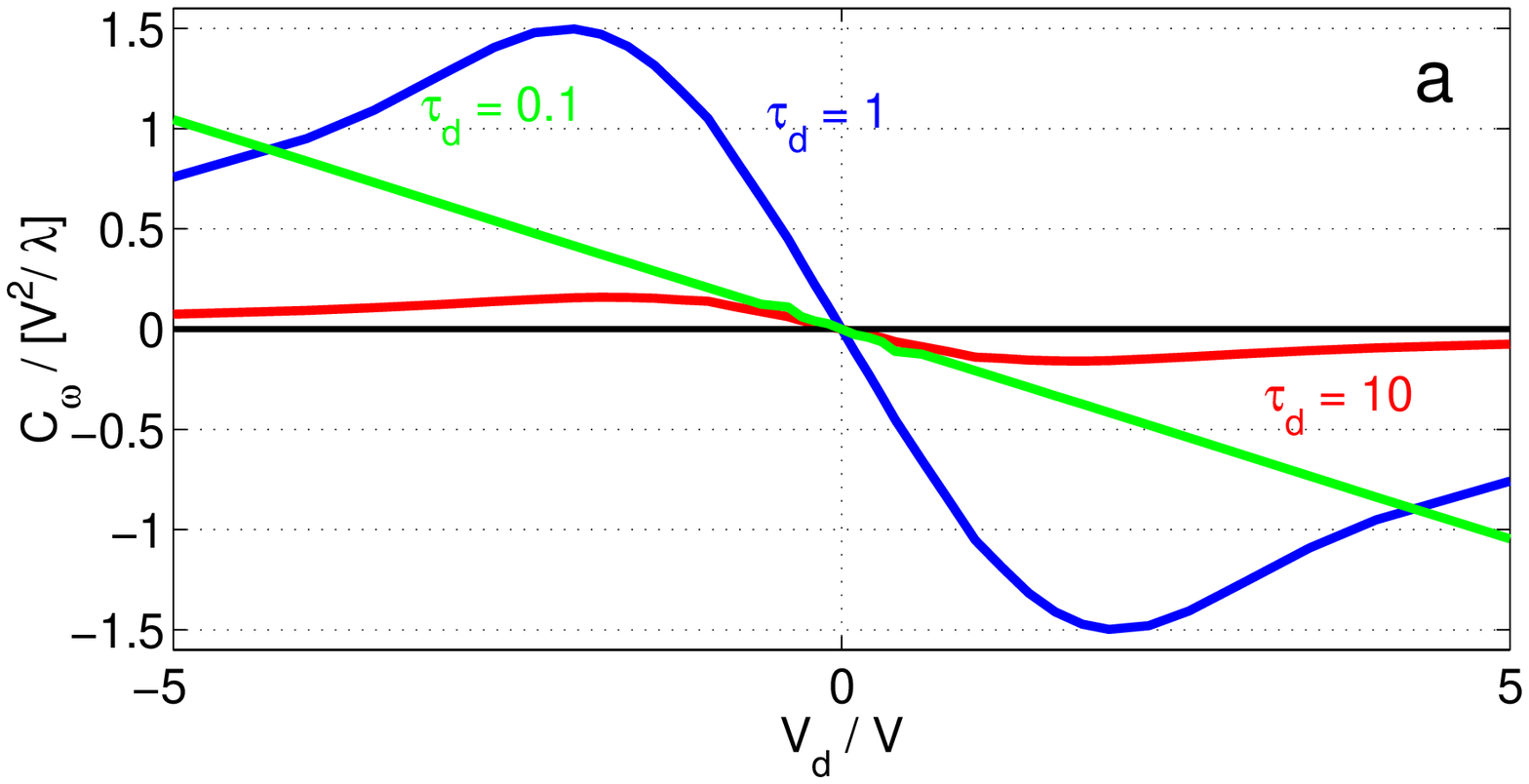}} \centerline{%
\includegraphics[height=4.2cm]{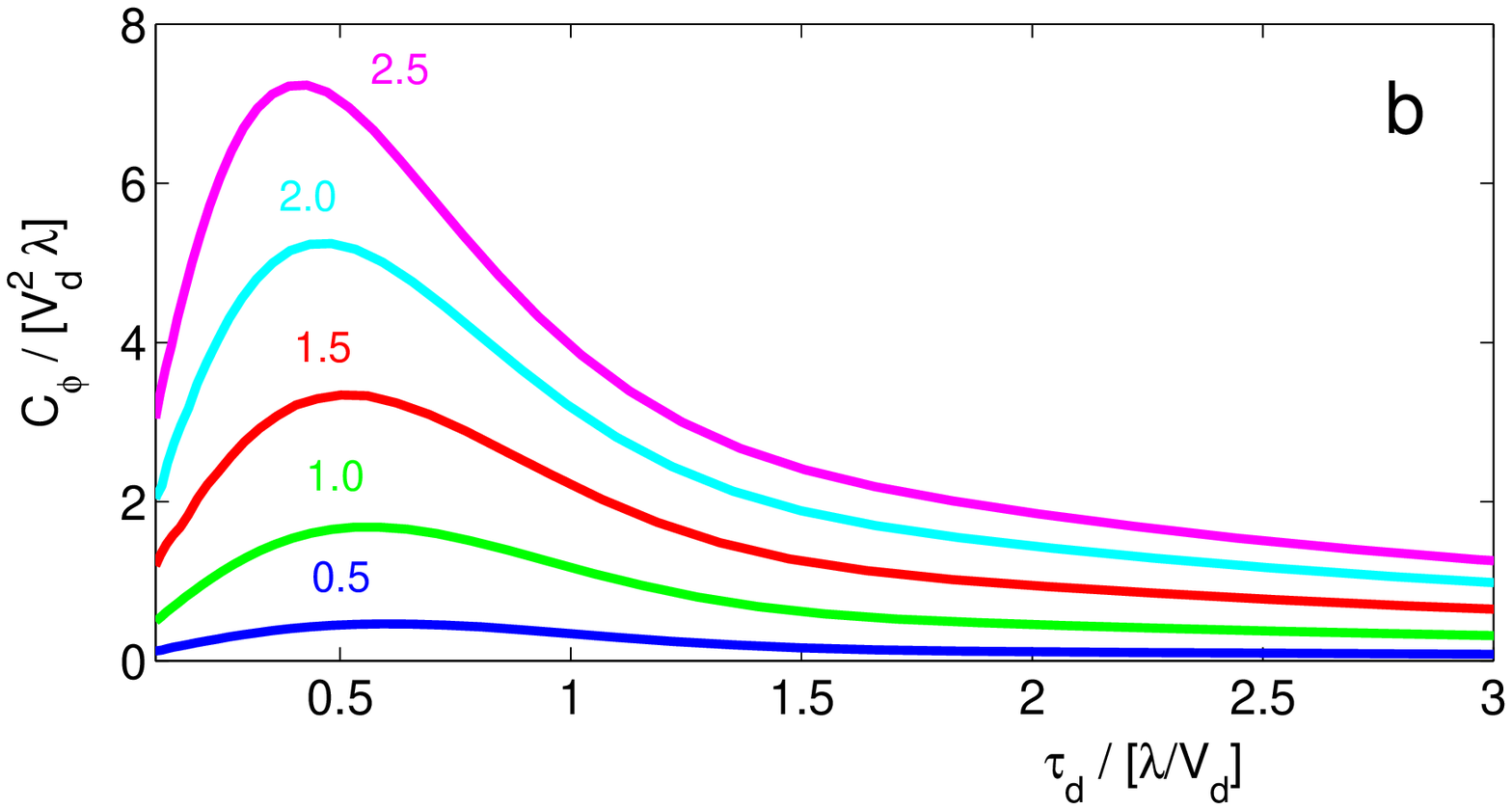}} 
\caption{The Lagrangian correlations $C_{\protect\omega}$ and $C_{\protect%
\phi}$ produced by the HDs. a) $C_{\protect\omega}$ as functions of the
average velocity $V_d$ for the values of the correlation time that label the
curves and $V=1.$ b) $C_{\protect\phi}$ as function of $\protect\tau_d$ for
the amplitudes $V$ of the stochastic velocity field that label the curves
and $V_d=4.$ }
\label{Figure 2}
\end{figure}

These correlations determine strange turbulent fluxes (STFs) of potential
and of vorticity. They consists of ordered motion of the fluctuations in
directions that are associated to their sign. This means that both the peaks
and the holes contribute to the flux. We show below that the HDs and the
STFs have strong nonlinear effects on turbulence evolution, which are
completely different of those induced by the random turbulent fluxes.

\section{Effects of HDs and STFs in fluid turbulence}

Turbulence relaxation in 2D ideal fluids is described by the
Euler equation with a stochastic initial condition

\begin{equation}
\partial _{t}\omega +\mathbf{v}\cdot \mathbf{\nabla }\omega =0,\ \ \mathbf{v}%
=-\mathbf{\nabla }\phi \times \mathbf{e}_{z}+V_{d}\mathbf{e}_{2},
\label{Euler}
\end{equation}%
where $\omega =\omega _{0}+\delta \omega $ is the vorticity, $\phi $ is the
stream function and $\mathbf{v}$ is the fluid velocity. The average velocity 
$V_{d}$ is a large scale nonuniform motion $V_{d}(x_{1})=V_{d}\exp
(-x_{1}^{2}/L^{2}),$\ where $L\gg \lambda _{1}.$\ It determines an average
(initial) vorticity $\omega _{0}(x_{1})=-2V_{d}~x_{1}\exp
(-x_{1}^{2}/L^{2})/L^{2}$ that adds to the fluctuation $\delta \omega .$
Thus, vorticity elements are advected by the velocity field and the
vorticity is conserved along the trajectories (\ref{eqm}). The nonlinearity
of the process is determined by the relations between $\mathbf{v}$ and $%
\omega .$

Two coupled equations for $\omega _{0}$ and for the amplitude of \
fluctuations $\left\langle \delta \omega ^{2}\right\rangle $\ are obtained
from (\ref{Euler}). The average equation extracted from (\ref{Euler})\
determines an equation for $\delta \omega ,$\ which can be formally solved
by integration along the trajectories (\ref{eqm}). Multiplying this solution
with $\delta \omega (\mathbf{x},t)$\ and with $\mathbf{v}(\mathbf{x},t)$
respectively, and averaging over the trajectories, one obtains

\begin{eqnarray}
\partial _{t}\left\langle \delta \omega ^{2}\right\rangle &=&-C_{\delta
\omega }\partial _{1}\omega _{0},  \label{omfluct} \\
\partial _{t}\omega _{0}+\mathbf{\nabla \cdot }\left\langle \mathbf{v~}%
\delta \omega \right\rangle &=&0  \label{omav}
\end{eqnarray}%
The turbulent flux of the vorticity is 
\begin{equation}
\left\langle \mathbf{v~}\delta \omega \right\rangle =-D_{1}\partial
_{1}\omega _{0}+C_{\delta \omega },  \label{flux}
\end{equation}%
where $D_{i}$\ are the diffusion coefficients obtained from the trajectories
(\ref{eqm}), $\partial _{t}$\ is the explicit time derivative and $\partial
_{1}$ is the partial derivative to $x_{1}$.

The correlation $C_{\delta \omega }$\ \ (\ref{Corvort}) shown in Figure 2a
appears in both equations. It determines a STF that consists of ordered
motion of the vorticity fluctuations in directions that have the sign of $%
-\delta \omega .$ The main effects of the HDs on turbulence can be deduced
from Eq. (\ref{omfluct})-(\ref{flux}). These equations are nonlinear due to $%
C_{\delta \omega }$ and $D_{1},$ which are rather complicated functions of
the parameters of the stochastic stream function, and also depend on time and on space (through $V_{d}$).

The amplitude of the vorticity fluctuations is not conserved, but it decays,
as seen in equation (\ref{omfluct}) where both $C_{\delta \omega }$ and $%
\partial _{1}\omega _{0}$\ are negative. The decrease of $\left\langle
\delta \omega ^{2}\right\rangle $ is the result of the ordered motion
combined with the conservation of the total vorticity $\omega =\omega
_{0}+\delta \omega $ along the trajectories. In the case of a completely
random motion, the change of the Lagrangian vorticity $\delta \omega
(x_{1}(t))$ is random and it does not influence the amplitude of the
fluctuations.

The equation of conservation of the average velocity (\ref{omav}) contains
the flux (\ref{flux}), which has a diffusive term and the contribution of
the HDs. The latter is not of advection type, but it actually determines a
source term%
\begin{equation}
\partial _{t}\omega _{0}+\partial _{1}\left( D_{1}\partial _{1}\omega
_{0}\right) =-(\partial C_{\delta \omega }/\partial V_{d})\omega _{0},
\label{omav2}
\end{equation}%
where the space derivative of $C_{\delta \omega }$\ was replaced by $%
\partial _{1}C_{\delta \omega }=(\partial C_{\delta \omega }/\partial
V_{d})\omega _{0}.$ The source term has the sign of $\omega _{0}$\ because $%
\partial C_{\delta \omega }/\partial V_{d}<0$\ for $V_{d}\lesssim V$ (Figure
2a). It determines the increase of $\left\vert \omega _{0}\right\vert .$ The
process is due to the STF, which brings positive fluctuations toward the
maximum of $\omega _{0}$ and negative $\delta \omega $ toward the minimum of 
$\omega _{0}.$

We note that Eqs. (\ref{omfluct}), (\ref{omav2}) represent a minimal frame
for a general analysis of the effect of the HDs. A more complicated equation
for the correlation of the vorticity fluctuations has to be added to the
system in order to describe the evolution of the initial turbulent state.
However, the importance of the HDs in the amplification of the average flow
can be\ quantitatively evaluated based on the present simplified
description. The correlation $C_{\delta \omega }$ is an increasing function
of the amplitude $V$\ of the stochastic velocity. This imposes a stochastic
initial condition with large amplitude of the vorticity fluctuation such
that $V\gg V_{d}.$\ In these condition $\partial C_{\delta \omega }/\partial
V_{d}<0$ even at the large values of $\tau _{d}$ that are expected for
turbulence evolution. The diffusion coefficient also increases with the
increase of $V,$ but slower that $C_{\delta \omega },$ and it decays at
large $\tau _{d}.$\ It means that, in these conditions, the process is
dominated by the effects of the HDs, which establish the characteristic time
scale of turbulence evolution.\ 

Thus, we have shown that the HDs and the STF of vorticity can determine a statistically relevant process of built up of large scale coherent motion at the expense of the small scale fluctuations. The basic individual process is similar to the evolution of small scale vortices on nonlinear sheared velocity \cite{Dubin}. The dynamics of the average vorticity in a turbulent environment described by Eqs. (\ref{omfluct}), (\ref{omav2}) shows that the separation of the vorticity elements according to their signs can contribute to the inverse cascade of energy that characterizes 2D fluid turbulence.

\section{Effects of HDs and STFs in turbulent plasmas}

We consider a plasma confined by a uniform magnetic field $B$\ taken along
the $\mathbf{e}_{3}$\ axis. A density gradient (along $\mathbf{e}_{1}$\
axis, with characteristic length $L_{n})$ makes plasma unstable \cite{GR}.
The drift wave instability produced by the electron kinetic effect and ion
polarization drift velocity is analyzed here using the\ method of test modes
on turbulent plasmas \cite{Vlad13}, \cite{MGS}. This method is based on the
separation of the distribution function into an approximate equilibrium $%
f_{0}$ and the response $h$ to the small perturbation of the potential (with
wave number components $k_{i}$ and frequency $\omega ),$ $\delta \phi \exp
(ik_{i}x_{i}-i\omega t),$\ that adds to the background potential $\phi _{b}(%
\mathbf{x},t).$\ The function $f_{0}(\mathbf{x},t)$ is the solution of the
approximate evolution equation obtained by neglecting the small terms. The
latter are not important at short time, but only at long times when the
small effects accumulate. The response $h$ can be linearized in the small
perturbation $\delta \phi .$ The solution of the dispersion relation yields
the frequencies $\omega (\mathbf{k})$ that can be supported by the system
and the tendency of amplification or damping given by the growth rate $%
\gamma (\mathbf{k}),$ the imaginary part of $\omega (\mathbf{k})$.\ \ These
quantities, which include the effects of the small terms neglected in the
evolution of $f_{0},$\ provide the short time change of the spectrum of the
background turbulence.\ 

Drift modes have small parallel wave numbers $k_{3}\ll k_{1},k_{2}$ and
small frequencies. The fast parallel motion of the electrons leads to the
adiabatic approximate equilibrium of the electrons and to the response to $%
\delta \phi $ 
\begin{equation}
\delta n^{e}=n_{0}(x)\frac{e\delta \phi }{T_{e}}\left( 1+i\sqrt{\frac{\pi }{2%
}}\frac{\omega -k_{y}V_{\ast }}{\left\vert k_{z}\right\vert v_{Te}}\right) 
\label{edn}
\end{equation}%
that does not depend on the background turbulence \cite{GR}. $n_{0}(x)$ is
plasma density, $T_{e}$ is the temperature of the electrons, $v_{Te}$ is
their thermal velocity, $V_{\ast }=\rho _{s}c_{s}/L_{n}$ is the electron
diamagnetic velocity,$\ \rho _{s}=c_{s}/\Omega _{i},$ $c_{s}=\sqrt{%
T_{e}/m_{i}}$ is the sound velocity and $\Omega _{i}=eB/m_{i}$ is the ion
cyclotron frequency. Using the constraint of neutrality one obtains the
short time equilibrium solution for the ions

\begin{equation}
f_{0}^{i}=n_{0}(x)F_{M}^{i}\exp \left( \frac{e\phi _{b}(\mathbf{x-V}_{\ast
}t)}{T_{e}}\right) ,  \label{f0i}
\end{equation}%
where $F_{M}^{i}$ is the Maxwell distribution of the ion velocities. Since $%
e\phi _{b}/T_{e}\ll 1,$ this distribution correspond to adiabatic density
fluctuations $\delta n/n\cong e\phi _{b}/T_{e}.$

The formal solution of the linearized equation for the response $h$ is

\begin{equation}
h(\mathbf{x},v,t)=-n_{0}(x)F_{M}^{i}\frac{e\delta \phi }{T_{e}}\left(
k_{2}V_{\ast }-\omega \rho _{s}^{2}k_{\perp }^{2}\right) \overline{\Pi },
\label{hi}
\end{equation}%
where the propagator $\overline{\Pi }$ is

\begin{eqnarray}
\overline{\Pi } &=&i\int_{-\infty }^{t}d\tau \ M(\tau )\ \exp \left[
-i\omega \left( \tau -t\right) \right] ,  \label{piim} \\
M(\tau ) &\equiv &\left\langle \exp \left[ \frac{e\phi _{b}(\mathbf{x}(\tau
))}{T_{e}}+i\mathbf{k\cdot }\left( \mathbf{x}(\tau )-\mathbf{x}\right) %
\right] \right\rangle ,  \label{M}
\end{eqnarray}%
and the average $\left\langle {}\right\rangle $\ is on the trajectories
obtained from equation (\ref{eqm}) in the background potential $\phi _{b}$
by integration backward in time from the initial condition at time $t.$ The
nonlinear constraint of the drift turbulence evolution is the
quasi-neutrality condition $\delta n^{i}=\delta n^{e},$\ which yields the
dispersion relation for test modes in turbulent plasma

\begin{equation}
-\left( k_{y}V_{\ast }-\omega \rho _{s}^{2}k_{\perp }^{2}\right) \overline{%
\Pi }^{i}=1+i\sqrt{\frac{\pi }{2}}\frac{\omega -k_{y}V_{\ast }}{\left\vert
k_{z}\right\vert v_{Te}}.  \label{dr}
\end{equation}%
The effects of $\phi _{b}(\mathbf{x},t)$ appear in the function $M(\tau )$
of the propagator (\ref{piim}).

The HDs lead to the correlation of the potential with the displacements in
the average (\ref{M}). Using the stationarity of the background turbulence
and the fact that the "initial" condition for a trajectory (\ref{eqm}) can
be any of its points, one obtains

\begin{equation}
\left\langle \phi _{b}(\mathbf{x}(\tau ))~x_{1}(\tau )\right\rangle
=\left\langle \phi _{b}(\mathbf{0})~x_{1}(\tau )\right\rangle .  \label{cor}
\end{equation}%
At large time $t\gg \tau _{d},$ this correlation can be approximated by $%
C_{\phi _{b}}(\tau -t),$\ where $C_{\phi _{b}}$ is the correlation (\ref%
{Corfi}) shown in Figure 2b. Neglecting for simplicity the effects of the
quasi-coherent structures described in \cite{Vlad13}, the function $M$ is
approximated by

\begin{equation}
M(\tau )\equiv \left\langle \exp \left[ ik_{1}V_{\phi }(\tau -t)-\frac{1}{2}%
k_{i}^{2}D_{i}(\tau -t)\right] \right\rangle ,  \label{M-HD}
\end{equation}%
where 
\begin{equation}
V_{\phi }=eC_{\phi _{b}}/T_{e}  \label{Vfi}
\end{equation}
plays the role of an average velocity along $\mathbf{e}_{1}$ axis, but it is
determined by the STF of the potential fluctuations that corresponds to
ordered motion of the positive fluctuations of $\phi _{b}$ with positive
velocity and of symmetrical motion of the negative fluctuations.

The dispersion relation has the solution

\begin{eqnarray}
\omega &=&\frac{k_{2}V_{\ast }-k_{1}V_{\phi }}{1+k_{\perp }^{2}}  \label{om}
\\
\gamma &=&c\frac{\left( k_{2}V_{\ast }+k_{1}V_{\phi }\right) \left(
k_{2}V_{\ast }+k_{1}V_{\phi }\right) k_{\perp }^{2}}{\left( 1+k_{\perp
}^{2}\right) ^{3}}-k_{i}^{2}D_{i},  \label{gam}
\end{eqnarray}%
where $c=\sqrt{\pi /2}/\left\vert k_{3}\right\vert v_{T_{e}}$ and $k_{\perp
}^{2}=k_{1}^{2}+k_{2}^{2}$.\ 

The velocity $V_{\phi }$ modifies both the frequency and the growth rate of
the test modes.

The new term in the frequency (\ref{om}) determines the drift of the
potential fluctuations along $\mathbf{e}_{1}$\ axis that is of STF type.
This potential flux drives STFs for the density fluctuations (due to the
adiabatic condition) and for the vorticity fluctuations (due to the
correlation (\ref{EC_fiom})). These fluxes have opposite signs, which depend
on the sign of the average velocity $V_{d}.$ The density STF determines
improvement or degradation of the confinement, depending on the values of $%
V_{d}$\ and on turbulence characteristics. The vorticity STF influences the
large scale vorticity that corresponds to the sheared rotation of the plasma 
\cite{Cad}.

The velocity $V_{\phi }$\ also modifies the growth rate (\ref{gam}). It
destabilizes a new type of modes that have $k_{2}=0,$ the zonal flow modes.
Their growth rate, obtained from equation (\ref{gam}) for $k_{2}=0,$ is not
zero due to $V_{\phi }$.\ This is a new mechanism for the generation of the
zonal flow modes. It is a nonlinear process determined by the HDs, which
yield the average velocity $V_{\phi }$\ in the propagator (\ref{piim}) that
drives oscillations of the potential along the $\mathbf{e}_{1}$ axis. A
significant effect appears when turbulence amplitude reaches high levels
such that $V_{\phi }$\ is comparable with $V_{\ast }.$\ 

A first evaluation of the importance of the process identified here can be
done by comparison\ with the results obtained in \cite{Vlad13}, \cite{VS17}
for the drift turbulence driven by the polarization drift. The latter
determines by correlation with the displacements a similar average velocity
in the propagator, which is much smaller that $V_{\phi }.$\ Despite this,\
the zonal flow modes can reach (in the case of a strong drive of the
instability) amplitudes that are high enough to produce the decay of the
turbulence \cite{VS17}. The HDs are more efficient for the generation of
zonal flow modes, because $V_{\phi }$\ is significantly larger (values of
the order $V_{\ast }$\ correspond to normalized amplitudes $e\Phi
/T_{e}\gtrsim 10^{-2}).$\ 

The processes studied in \cite{Vlad13}, \cite{VS17} (correlations of the
polarization drift, effects of the ion structures and flows produced by
trajectory trapping) are neglected here. Also, the Reynolds stress does not
play an important role in these models. As a consequence, one can conclude
that the effects of the HDs add to the above mechanisms. Synergistic
influences are expected, especially through the combination of different
contributions to the total average velocity along $\mathbf{e}_{1}$ axis.

\section{Discussion and conclusions}

The hidden drifts (HDs) are found in the statistics of the trajectories (\ref%
{eqm}) in 2D incompressible stochastic velocity fields in the
presence of an average velocity $V_{d}\mathbf{e}_{2}.$ They are ordered
components of the motion oriented perpendicular to the average velocity $%
V_{d}\mathbf{e}_{2},$ which exactly compensate one another and do not yield
average displacements of the trajectories. Direct effects of the HDs on test
particle transport do not appear in this case. In spite of this, generation
of convective transport through the perturbation of the symmetry of the HDs
is expected in more complex models that include a small compressibility of
the velocity field. A first process of this type \cite{VS2018} shows that
the convection produced by the HDs can be comparable or larger than the
diffusive transport.

The most important effect of the HDs is connected to the Lagrangian
correlations that they determine. A special type of fluxes are generated,
which consist of displacements of the positive and negative fluctuations in
opposite directions. This is a rather unexpected behavior since the equation
for the trajectories (\ref{eqm}) does not depend on the advected quantity.
For this reason we have denoted these correlations strange turbulent fluxes
(STFs). They are due to the pair of HDs that have the orientation related to
the sign of the initial potential or vorticity.

The HDs and the STF of potential and vorticity fluctuations were determined
using the decorrelation trajectory method as functions of the parameters of
the test particle transport model. The HDs are essentially produced by the
average velocity $V_{d}\mathbf{e}_{2},$ which modifies the structure of the
contour lines of the total potential and the statistics of the velocity
along these lines.\ \ \ 

We have discussed the effects of the STFs in two cases: the relaxation of
turbulent states in 2D ideal (inviscid) fluids and drift
turbulence in collisionless (hot) plasmas confined in strong magnetic
fields. These physical systems are described by similar evolution equations
that represent the advection along the characteristics obtained from Eq. (%
\ref{eqm}). They are formally linear, but there are in both cases nonlinear
constraints that strongly influence turbulence evolution. The nature of the
nonlinearity and the physical significance of the advected fields are
completely different in the two systems. We have shown that these
differences lead to completely different effects of the HDs, although they
are the same in the two systems.

In the process of relaxation of turbulent states in 2D ideal
fluids, the STF of the vorticity fluctuations determines the increase of the
large scale vorticity and velocity, accompanied by the decrease of the
amplitude of the small scale vorticity fluctuations.

In magnetically confined plasmas, the approximate adiabatic response that
characterizes drift-type turbulence determines STFs of potential and
vorticity fluctuations. They modify the frequencies of the test modes on
turbulent plasmas and generate zonal flow modes.

{\bf Acknowledgments}

This work has been carried out within the framework of the EUROfusion
Consortium and has received funding from the Euratom research and training
programme 2014-2018 under grant agreement No 633053 and also from the
Romanian Ministry of Research and Innovation. The views and opinions
expressed herein do not necessarily reflect those of the European Commission.

\end{document}